\documentclass[preprint,aps,nofootinbib,tightenlines]{revtex4}

\usepackage{graphicx}

\usepackage{amsmath}
\usepackage{amsfonts}
\usepackage{amssymb}

\newcommand{\be}{\begin{equation}}
\newcommand{\ee}{\end{equation}}
\newcommand{\bear}{\begin{eqnarray}}
\newcommand{\eear}{\end{eqnarray}}
\newcommand{\beqstar}{\begin{eqnarray*}}
\newcommand{\eeqstar}{\end{eqnarray*}}

\newcommand{\ov}[1]{\overline{#1}}

\begin{document}


\vspace*{1.5cm} 
\title{Little Hierarchy, Little Higgses, and a Little Symmetry\vspace{0.5cm}}

\author{Hsin-Chia Cheng and Ian Low}
\affiliation{Jefferson Physical Laboratory,
Harvard University, Cambridge, MA 02138
\vspace*{0.5cm}}

\begin{abstract}
\vspace*{0.5cm} 

Little Higgs theories are an attempt to address the  ``little
hierarchy problem,'' {\it i.e.},  the tension between the naturalness
of the electroweak scale and the precision electroweak measurements
showing no evidence for new physics up to 5 -- 10 TeV.  In little
Higgs theories, the Higgs mass-squareds are protected at one-loop
order from the quadratic divergences. This allows the cutoff of the
theory to be raised up to $\sim 10$ TeV, beyond the scales probed by
the current precision data.  However, strong constraints can still
arise from the contributions of the new TeV scale particles which
cancel the one-loop quadratic divergences from  the standard model
fields, and hence re-introduces the fine-tuning problem.  In this
paper we show that a new symmetry, denoted as $T$-parity, under which
all heavy gauge bosons and scalar triplets are odd,  can remove all
the tree-level contributions to the electroweak observables and
therefore makes the little Higgs theories completely natural.   The
$T$-parity can be manifestly implemented in a majority of little Higgs
models by following the most general construction of the low energy
effective theory \`a la Callan, Coleman, Wess and Zumino.   In
particular, we discuss in detail how to implement the $T$-parity in
the littlest Higgs model based on $SU(5)/SO(5)$.  The symmetry
breaking scale $f$ can be even lower than 500 GeV if the contributions
from the higher dimensional operators due to the unknown UV physics at
the cutoff are somewhat small.  The existence of $T$-parity has
drastic impacts on the phenomenology of the little Higgs theories. The
$T$-odd particles need to be pair-produced and will cascade down to
the lightest $T$-odd particle (LTP) which is stable.  A neutral LTP
gives rise to missing energy signals at the colliders which can mimic
supersymmetry. It can also serve as a good dark matter candidate.

\end{abstract}


\maketitle

\section{Introduction}
\label{sec:introduction}

The origin of the electroweak symmetry breaking, a fundamental
mystery of the weak interactions, will be probed directly by high
energy experiments in the coming decade. In the standard model (SM) 
the symmetry breaking is triggered by the vacuum expectation value (VEV)
of a scalar Higgs field. 
The quantum
correction to the mass of the Higgs particle is, however, very sensitive to
the ultraviolet (UV) physics. To produce the observed electroweak breaking
scale, the radiative corrections to the Higgs mass-squared 
need to be fine-tuned, 
unless new particles are introduced at around the 1 TeV scale
to cut off the quadratically divergent contributions.
Generally speaking, if the scale of the electroweak symmetry breaking is
to be
stabilized naturally, we expect new physics to show up at the scale $\sim$
1 TeV or below.

At energies below the scale of new physics, new particles can be
integrated out to obtain a set of higher dimensional
operators involving the standard model fields only~\cite{Weinberg:1978kz}.
These higher dimensional operators can contribute to experimental
observables even at energies beneath the scales of new physics, 
and the size of these operators are constrained by
precision measurements~\cite{Buchmuller:1985jz}. 
The strongest bounds are on the
operators violating the (approximate) symmetries of the standard model, such
as baryon number, flavor and CP symmetries. The dimensionful parameters
suppressing these higher dimensional operators typically need to be
above 100 TeV due to the absence or rareness of events which violate these
symmetries. 
This implies that the new physics at the TeV scale should also respect
these symmetries (at least approximately).
On the other hand, operators preserving the symmetries of the
standard model are expected to be induced by new physics at the TeV
scale. These operators can be probed by the electroweak precision measurements.
The advances of the electroweak precision measurements 
in the past decade put
significant constraints on many of these operators. The most constrained
operators need to be suppressed by the energy scales 
$5 - 10$ TeV~\cite{Barbieri:1999tm}, assuming that all dimensionless
coefficients are ${\cal O}(1)$. Taken at face value, it seems to 
indicate that there is no new physics up to $\sim 10$ TeV, which creates
a tension, known as the little hierarchy problem, 
with the naturalness requirement that new physics should appear at $\sim 1$ TeV or below
in order for the electroweak symmetry breaking to be stabilized.
This discrepancy might be accidental and the electroweak
scale is fine-tuned. However, it is also quite possible
that the little hierarchy problem is in fact giving us an important hint of
the new physics at the TeV scale and ought to be taken more seriously.

In the supersymmetric extension of the standard model,
a conserved $R$-parity can be imposed to eliminate all the 
tree-level contributions to the electroweak observables from the 
superpartners so that the coefficients of the higher dimensional operators
are naturally suppressed by a loop factor. This would relieve 
the little hierarchy problem.
However, there is still a
related fine-tuning problem in supersymmetric theories. In the standard
model the Higgs particle has a special status as the only scalar in the theory.
In supersymmetric theories there are, nevertheless,
many superpartners which are also scalars. Generically superpartner
masses are expected to be in the same order as the Higgs mass since the
Higgs is not special. 
However, the loop contributions to the electroweak
observables from some superpartners as light as ${\cal O}$(100 GeV)
could still be dangerous ({\it e.g.}, $\Delta \rho$ from top squark loops).
Moreover, in the minimal supersymmetric standard model (MSSM), raising
the lightest Higgs mass above the experimental bound requires large
top squark masses.
On the other hand, 
the radiative corrections to the Higgs mass from the top squark masses
are at least as large as the top squark masses themselves
if one requires the
theory to be valid up to the energy scale of the gauge coupling
unification. As a result, such
heavy top squarks, though more phenomenologically 
desirable from above discussion, introduce fine-tuning at the level
of a few percent or less.
This naturalness problem in supersymmetry has been discussed in many places
before~\cite{Giusti:1998gz,Batra:2003nj,Harnik:2003rs,Barbieri:2003dd,nimatalk}.
The fact that so far we have found neither  
the Higgs, the superpartners, nor deviations from the electroweak precision
data, indicates that the electroweak scale still appears to be 
somewhat fine-tuned for the simplest supersymmetric extensions of the standard
model, even if supersymmetry is the answer to the physics
beyond the standard model.

Little Higgs
theories~\cite{Arkani-Hamed:2001nc,Arkani-Hamed:2002pa,Arkani-Hamed:2002qx,Arkani-Hamed:2002qy,Gregoire:2002ra,Low:2002ws,Kaplan:2003uc,Chang:2003un,Chang:2003zn,Skiba:2003yf} 
provide an alternative approach to 
stabilize the electroweak scale based on a non-linear chiral Lagrangian
of a coset space $G/H$. The Higgs is light because it is a 
pseudo-Nambu-Goldstone boson (PNGB) with the unique property that its mass 
is protected at one-loop order from the quadratic divergences. 
The cutoff can therefore
be raised above $\sim 10$~TeV, beyond the scales probed by
the current electroweak precision measurements. Consequently 
the contributions from
the UV physics above the cutoff are in general safe from the
electroweak constraints, as long as the unbroken group $H$ contains a custodial
$SU(2)$ symmetry. The one-loop quadratically divergent contributions
to the Higgs mass from the SM top quark, gauge fields and Higgs itself
are cancelled by contributions from 
new particles at the $\sim 1$ TeV scale with
the same spins as the SM particles. 
In contrast to supersymmetry, there is a natural separation in the mass scale of the Higgs and 
the new TeV particles, since the Higgs is the only PNGB whose
mass is suppressed by a two-loop factor. 

The electroweak observables also receive contributions from 
the new TeV particles in little Higgs theories, and 
there have been
several studies on this issue for various little Higgs 
models~\cite{Hewett:2002px,Csaki:2002qg,Csaki:2003si,Gregoire:2003kr,Kilic:2003mq,Chen:2003fm,Casalbuoni:2003ft,Kilian:2003xt,
  Yue:2004xt}.
The results are quite model and parameter dependent, although the general
impression is that the electroweak precision data impose strong constraints
on the symmetry breaking scale of $G$ in most of the parameter space, 
and hence the TeV particles which cancel the quadratic divergences are
required to be heavy. This in turn
re-introduces fine-tuning into the little Higgs models.
A closer look shows that the strongest constraints
come from the direct couplings of the SM fields to the new gauge bosons,
as well as the VEV of any $SU(2)$ scalar triplet which arises from coupling
to the SM Higgs. These couplings, however, are not an essential part of 
the little Higgs theories; they do not participate in the cancellation 
of the quadratic
divergences of the Higgs mass corrections. If there is a natural way,
{\it e.g.}, by imposing a new symmetry, 
to suppress these dangerous couplings in 
a little Higgs model, it would be completely consistent
with the precision electroweak constraints without any fine tuning.
Indeed this was achieved in Ref.~\cite{Cheng:2003ju} for a three-site
moose model, where a $T$-parity forbids all these dangerous couplings.

In this paper, we discuss how the dangerous couplings in little Higgs theories can be
removed in a more general context. 
The $T$-parity is usually easy to be incorporated into the scalar and the gauge
sectors. However, it was not respected by the fermion sectors in the conventional
little Higgs models. This is due to the fact that 
in the original models and the follow-up phenomenological studies,  
the SM fermions are assigned to particular
representations under the broken group $G$. 
{}From a low energy effective theory point of view, however, only  
the symmetry transformation property of a particle under
$H$ is unambiguous since only the unbroken group $H$ is manifest.
As Callan, Coleman, Wess and Zumino~\cite{Coleman:sm,Callan:sn} (CCWZ)
showed long time ago, by starting with a
Lagrangian manifestly invariant under the unbroken group $H$ and matter
content furnishing linear representations of $H$, one can construct a 
Lagrangian non-linearly realizing the full symmetry $G$ with the help of
Nambu-Goldstone fields which parametrize the coset space $G/H$. 
From the low energy effective theory perspective, this is a 
more appropriate way to construct the effective Lagrangian and it is not necessary to
introduce linear representations of the full group $G$ at all.   
The advantage of this approach in our context is that there is a natural separation 
of the model-dependent
dangerous interactions from those essential interactions 
which engineer the little Higgs mechanism.
In a majority of the little Higgs models, one can easily identify a little
symmetry---the $T$-parity, which can be imposed to remove all the
unwanted dangerous couplings, rendering these models completely natural.

In the next section, we start with the minimal moose model of 
Ref.~\cite{Arkani-Hamed:2002qx}. It is an ideal example for the
illustrative purpose because of its similarity to the very familiar
QCD chiral Lagrangian. In section~\ref{sec:coset} we move on to the
most popular $SU(5)/SO(5)$ littlest Higgs model~\cite{Arkani-Hamed:2002qy}
and show how the $T$-parity can be implemented to remove the strong 
constraints from the electroweak precision data. The existence of the
$T$-parity has a drastic impact on the little Higgs phenomenology, which
will be briefly discussed in section~\ref{sec:pheno}. We will
comment on general models in which $T$-parity can or cannot 
be naturally implemented and draw the conclusions in section~\ref{sec:con}.

\section{An $SU(3)$ minimal moose model with $T$ parity}
\label{sec:twosite}

We start our discussion with the minimal moose
model~\cite{Arkani-Hamed:2002qx}, a two-site model based on $SU(3)$
global symmetry. This model does not have a custodial $SU(2)$ symmetry, and 
a large contribution to the $\Delta \rho$ parameter is already present in the 
non-linear sigma model, which is independent of
the contributions from the TeV scale
particles. However, this can be easily cured
by replacing the $SU(3)$ group with another group, such as $SO(5)$,  
containing a custodial $SU(2)$. 
Despite of the strong electroweak constraints on this model, we will
nevertheless use it to illustrate our main 
point because the model shares the same global symmetry breaking pattern, 
$SU(3)_L\times SU(3)_R \to SU(3)_V$, with the very familiar QCD chiral
Lagrangian. The purpose here is to show how 
a $T$-parity can be naturally imposed 
to get rid of the dangerous tree-level couplings
between SM fermions and the heavy gauge bosons.

The QCD chiral Lagrangian has a well-known $Z_2$ symmetry: the parity which
exchanges the chirality $L \leftrightarrow R$. Baryons, transforming as
octets under the unbroken $SU(3)_V$, can be incorporated in the chiral
Lagrangian in many different ways. The freedom lies in the assignments of
the full $SU(3)_L\times SU(3)_R$ representations under which baryons 
transform, since there are many representations containing an octet of the
unbroken $SU(3)_V$ after chiral symmetry breaking. Not every assignment
respects the parity symmetry in a manifest way. Alternatively,
we could also not worry about how baryons transform under the full chiral
group and contend ourselves with the fact that baryons form an octet under
the unbroken group \`a la CCWZ.
In any case, parity is a good symmetry of QCD and 
can be imposed in the chiral
Lagrangian~\cite{Manohar:1983md,georgibook}; the parity transformation is
non-linear and involves the Goldstone bosons if baryons are assigned to a
parity non-invariant representation of $SU(3)_L\times SU(3)_R$.

We run into a completely similar situation in the minimal moose
model. There is an incarnation of the parity in QCD: the $Z_2$ reflection
which interchanges the two sites. 
However, a na\"{i}ve assignment of the SM fermion
under the full symmetry group does not respect the reflection symmetry, resulting in
direct couplings between the standard model fermions and the heavy
gauge bosons. Nevertheless, as in the QCD chiral Lagrangian, a linear $Z_2$ parity
can be consistently implemented to forbid these unwanted couplings.
The parallel problem in the QCD is how to write down a chiral Lagrangian
including fermions which linearly realizes the parity. This problem has been
explicitly worked out in Ref.~\cite{Manohar:1983md,georgibook}, 
which is an adaption
of CCWZ to the chiral symmetry breaking. We will see that, by
following this approach, a $T$-parity which removes the electroweak
constraints from the new TeV particles in the minimal moose model
will become manifest.

The minimal moose model is a two-site $SU(3)$ model with four links $X_i =
\exp(2 i x_i/f), i=1,...,4$. We gauge the same $SU(2)\times U(1)$ subgroup
within both $SU(3)$ groups with equal strength.\footnote{This is slightly different from
the original model which gauged an $SU(3)$ on one site and $SU(2)\times
U(1)$ on the other. The reason is we want the gauge sector to respect the
$Z_2$ reflection symmetry. This minor difference does not affect the little Higgs mechanism.} 
The Goldstone boson
matrices, $x_i$, each contains a triplet, a doublet and a singlet under
the unbroken diagonal $SU(2)_V$ gauge group, 
which we take to be the electroweak
$SU(2)_W$. This model has a large global symmetry $[SU(3)_L\times SU(3)_R]^4$
spontaneously broken down to $[SU(3)_V]^4$, with the non-linear realization
\begin{equation}
 X_i \to L_i X_i R_i^\dagger, \quad i=1,...,4.
\end{equation}
The effective Lagrangian is written as
\begin{equation}
{\cal L} = {\cal L}_G + {\cal L}_X + {\cal L}_\psi + {\cal L}_t.
\end{equation}
Here ${\cal L}_G$ contains the conventional non-linear sigma model 
kinetic terms
and gauge interactions, whereas ${\cal L}_X$ includes the plaquette
operators which give rise to Higgs quartic interactions:
\begin{equation}
{\cal L}_X = \kappa f^4 \left[ \mbox{Tr}(X_1 X_2^\dagger X_3 X_4^\dagger) +
  \mbox{Tr}(X_2 X_3^\dagger X_4 X_1^\dagger) \right]+ {\rm h.c.},
\label{plaquette}
\end{equation}
where $f \sim 1$ TeV sets the cutoff of the theory to be at $\Lambda \sim
4\pi f \sim 10$ TeV. The gauge and scalar sectors obviously respect the 
parity,
\begin{eqnarray}
A_L &\leftrightarrow& A_R, \;\; \mbox{where $A_{L,R}$ are gauge fields 
for $[SU(2)\times U(1)]_{L,R}$,}\\
X_i &\leftrightarrow& X_i^\dagger ,
\end{eqnarray}
under which the SM gauge bosons are even, and all the Goldstone bosons and
heavy gauge bosons are odd. In order to make the standard model
Higgs even under the parity, we note that
the matrix $\Omega_X = \mbox{diag}(1,1,-1)$ commutes with all the gauge
generators and can be included in the parity transformation to
flip the parity of the Higgs doublet without affecting anything else.
(This is in fact equivalent to a hypercharge rotation.)
We therefore define the $T$-parity for the gauge and the scalar sectors as,
\begin{eqnarray}
A_L &\leftrightarrow& A_R, \\
X_i &\leftrightarrow& \Omega_X X_i^\dagger \Omega_X.
\end{eqnarray}
Under the $T$-parity, all SM gauge bosons and Higgs doublets are even, and
the heavy gauge bosons and scalar triplets and singlets are odd. 
One can immediately see that with the $T$-parity, 
there is no mixing between the SM gauge bosons and the heavy gauge bosons,
and the coupling of the
scalar triplets to two Higgs doublets is forbidden. So there is no
tadpole term to induce the triplet VEVs after electroweak symmetry breaking.

In the fermionic sector ${\cal L}_\psi$, the right-handed
(electroweak singlet) fermions can be easily incorporated to respect the
$T$-parity: we assign them to be charged only under the diagonal
$U(1)_V$, ({\it i.e.,} they carry equal charges under $U(1)_L$ and $U(1)_R$,)
\begin{equation}
{\cal L}_{\psi} \supset 
\ov{\psi}_S\, \sigma^\mu (\partial_\mu + i g_Y Y_{\psi_S} B_\mu)\, \psi_S
=\ov{\psi}_S\, \sigma^\mu \left(\partial_\mu + 
i \sqrt{2} g_Y \frac{Y_{\psi_S}}{2} (B_{L\mu}+B_{R\mu})\right) \psi_S,
\label{eq:singlet}
\end{equation}
where $\sigma^\mu=(1, \vec{\sigma})$ and 
$B_\mu$ is the SM hypercharge gauge field which is a linear combination
of the $U(1)_L$ gauge field $B_{L\mu}$ and $U(1)_R$ gauge field $B_{R\mu}$.
To realize the $T$-parity linearly for the left-handed (electroweak doublet)
fermions, we
define the matrix $\xi=\exp(i x /f)$, so that $\xi$ is the square root of
the link field, $\xi^2 = X$~\cite{Manohar:1983md,georgibook}. Under the
global $SU(3)_L\times SU(3)_R$ transformations, 
\begin{equation}
\xi \to \xi^\prime = L \xi U^\dagger = U \xi R^\dagger,
\end{equation}
where $U$ belongs to the unbroken $SU(3)_V$ and is a non-linear function of
$L, R,$ and the Goldstone bosons $x$. Now for each electroweak doublet
fermion in the standard model, $\psi_{D}$, we take it to transform under
$SU(3)_L \times SU(3)_R$ as
\begin{equation}
\psi = \left( \begin{array}{c}
                     \psi_{D} \\
                       0        
               \end{array}        \right)
 \to U \psi.
\label{assign}
\end{equation}
If $\psi$ were a complete multiplet of the unbroken $SU(3)_V$,
Eq.~(\ref{assign}) would define a
a non-linear representation of the $SU(3)_L \times SU(3)_R$ group.
To write down the kinetic term for the fermions, we need to compute the
Maurer-Cartan one form \cite{Coleman:sm,Callan:sn}
\begin{equation}
\xi^\dagger D_\mu \xi \equiv \xi^\dagger (\partial_\mu + i A_\mu^a Q_V^a + i
   {\cal A}_\mu^a Q_A^a ) \xi \equiv v_\mu^a T^a + p_\mu^a X^a,
\label{oneform}
\end{equation}
where $A_\mu^a, {\cal A}_\mu^a$ are the unbroken and broken gauge fields,
respectively, and $T^a, X^a$ are the unbroken and broken
generators. The Maurer-Cartan one form takes value in the Lie algebra of
the full symmetry group $G$, therefore it can be written as linear combinations
of $T^a$ and $X^a$. The components of $\xi ^\dagger D_\mu \xi$ in the
direction of unbroken and broken generators are defined as $v_\mu^a$ and
$p_\mu^a$ respectively. 
 Under the $U$ rotation, $v_\mu$ transforms like a gauge field
and $p_\mu$ transforms covariantly,
\begin{equation}
v_\mu^a T^a \to U v_\mu^a T^a U^\dagger + U (\partial _\mu U^\dagger),
\quad \quad p_\mu^a X^a \to U p_\mu^a X^a U^\dagger.
\end{equation}
These objects can be used to write down Lagrangians invariant under the
full broken group $G$. For example, a fermion kinetic term is given by
\begin{equation}
\ov{\psi}\, \bar{\sigma}^\mu (\partial_\mu + v_\mu^a T^a + i\,g_Y (Y_{\psi_D}-Y_H)
 B_\mu) \psi,
\label{fermionkinetic}
\end{equation}
where $\bar{\sigma}^\mu=(1, -\vec{\sigma})$. Apart from the gauge interactions
it realizes the full $SU(3)\times
SU(3)$ non-linearly. The additional term involving
$B_\mu$ serves to give the correct hypercharge to the SM fermion $\psi_D$. 

There is a compact way to write down
$v_\mu^a T^a$ by noting that, under the parity which interchanges $L
\leftrightarrow R$, the generators $T^a \to T^a$ and $X^a \to - X^a$.
Then Eq.~(\ref{oneform}) turns into
\begin{equation}
\xi D_\mu \xi^\dagger = \xi (\partial_\mu + i A_\mu^a Q_V^a - i
   {\cal A}_\mu^a Q_A^a ) \xi^\dagger = v_\mu^a T^a - p_\mu^a X^a.
\end{equation}
The fermion kinetic term, Eq.~(\ref{fermionkinetic}), can now be written as
\begin{equation}
\ov{\psi}\, \bar{\sigma}^\mu 
\left(\partial_\mu + \frac12(\xi^\dagger_4 D_\mu \xi_4 + \xi_4 D_\mu
\xi^\dagger_4)+ i\,g_Y (Y_{\psi_D}-Y_H)
 B_\mu \right) \psi,
\label{fermion}
\end{equation}
where, to be more concrete, we have inserted $\xi_4$, the square root of
the fourth link $X_4$, although it could be any of the links.
Eq.~(\ref{fermion}) is manifestly invariant under $T$-parity, since
the SM fermions are taken to be $T$-even. By expanding $v_\mu^a$,
one can check that at lowest order (with no Goldstone field), 
it only contains the standard model gauge interactions for the fermions, 
but not the direct couplings between the fermions and the heavy gauge bosons
${\cal A}_\mu$.  
 
Another term invariant under the full $SU(3)_L \times SU(3)_R$ symmetry
which we can write down is 
\begin{equation}
g_A \ov{\psi}\, \bar{\sigma}^\mu p_\mu^a X^a \psi= g_A \ov{\psi}\,
\bar{\sigma}^\mu (\xi^\dagger D_\mu \xi -\xi D_\mu \xi^\dagger) \psi,
\label{T-violating}
\end{equation}
with an {\em arbitrary} coefficient $g_A$. 
It contains the dangerous direct couplings
between the SM fermions and the heavy gauge bosons, which is the origin
of the strong electroweak precision constraints on the little Higgs models.
However, one can easily see that this term is $T$-odd, since the $T$-parity
interchanges $\xi^\dagger D_\mu \xi$ and $\xi D_\mu
\xi^\dagger$ in Eq.~(\ref{T-violating}),\footnote{Note that $\Omega_X Q_{A,V}^a \Omega_X = Q_{A,V}^a$,
and $\Omega_X \psi = \psi$ since $\psi$ only has 
the upper two components.} and can be forbidden 
by imposing the $T$-parity. The nice thing about following the approach of
Ref.~\cite{Coleman:sm,Callan:sn,Manohar:1983md,georgibook} is that the
essential $T$-preserving interactions and the model-dependent $T$-violating
interactions are naturally separated, thus allowing for an easy identification
of the $T$-parity which removes the strong electroweak constraints on the
little Higgs models.

In terms of gauge fields $A_\mu$ and ${\cal A}_\mu$,
Goldstone fields $x_i$, and the fermion $\psi$, the $T$-parity defined as
\begin{equation}
T: A_\mu \to A_\mu, \quad {\cal A}_\mu \to - {\cal A}_\mu, \quad x_i \to
-\Omega_X x_i \Omega_X, \quad \psi \to  \Omega_X \psi
\label{tparity}
\end{equation}
is an exact symmetry of ${\cal L}_G + {\cal L}_X + {\cal
L}_\psi$. The remaining task is to write down the standard
model Yukawa coupling ${\cal L}_t$, which was worked out in
Ref.~\cite{Arkani-Hamed:2002qx}, except that we need to $T$-symmetrize the
Yukawa interactions. For the top Yukawa coupling we introduce a 
vector-like pair of
$T$-even, colored Weyl fermions $u'$ and ${u'}^c$:
\begin{equation}  
{\cal L}_t \supset  \frac{\lambda}2 f (b\ t\ u^{\prime}) 
\left[\xi_1 \xi_2^\dagger + \Omega_X
 \xi_1^\dagger \xi_2 \Omega_X \right] 
         \left(   \begin{array}{c}
                          0  \\
                          0  \\
                          u_3^c
                  \end{array}         \right)   
    + \lambda^\prime f {u'}^c u'  ,
\label{top}
\end{equation}   
so that the quadratically divergent contribution to the Higgs mass-squared
induced by the large top Yukawa coupling is cancelled by the additional
fermion $u^\prime$. For the light up-type quarks, the Yukawa coupling has the same
form as Eq.~(\ref{top}) except that $u'$ and ${u'}^c$ are not needed, 
whereas for the
down-type quarks and charged leptons, the Yukawa coupling can be written as
\begin{equation}
{\cal L}_t \supset \frac{\lambda_d}2 f (0\ 0\ d^c) \left[\xi_1 \xi_2^\dagger + \Omega_X
 \xi_1^\dagger \xi_2 \Omega_X \right] 
    \left( \begin{array}{c}
                 {q}\\
                  0 
           \end{array}   \right)    +
    \frac{\lambda_e}2 f (0\ 0\ e^c) \left[\xi_1 \xi_2^\dagger + \Omega_X
     \xi_1^\dagger \xi_2 \Omega_X \right] 
    \left( \begin{array}{c}
                 {l} \\
                 0 
           \end{array}   \right) .
\end{equation}
With these Yukawa interactions, the $T$-parity defined in Eq.~(\ref{tparity})
is also an exact symmetry of the Yukawa sector ${\cal L}_t$, and hence a
symmetry of the full effective Lagrangian ${\cal L}$.

\begin{figure}[tb]
\includegraphics[width=6cm]{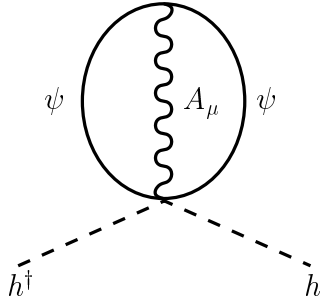}
\caption{\label{figureone}\it The two-loop diagram with the 
quartic divergence.}
\end{figure}

The matter content of the $T$-invariant $SU(3)$ minimal moose model is very
similar to the original minimal moose model in
Ref.~\cite{Arkani-Hamed:2002qx}. Because the gauge group is 
$[SU(2)\times U(1)]^2$, only two
scalar triplets and singlets are eaten through the Higgs mechanism after
symmetry breaking.  
Na\"ively it seems that in the $T$-invariant model there are three doublets
remaining light
since only one becomes heavy due to the plaquette operators Eq.~(\ref{plaquette}).
An additional doublet would have been eaten by the heavy gauge bosons if the
gauge group were $SU(3)\times SU(2)\times U(1)$ as in the original
minimal moose model.
However, the fermions in Eq.~(\ref{assign}) do not form a
complete multiplet of the global symmetry. The kinetic term
Eq.~(\ref{fermion}) breaks all chiral symmetries associated with $\xi_4$. 
The scalar
doublet residing in $\xi_4$ is no longer protected and picks up
a mass of the order of $gf \sim 1$ TeV. 
Diagrammatically, a two loop {\it quartic-divergent} contribution to 
the doublet mass-squared is indeed generated through the
two-loop diagram in Fig.~\ref{figureone},\footnote{This contribution was
first noted by Nima Arkani-Hamed.}
\begin{equation}
m_h^2 \sim \frac{g^2}{(16\pi^2)^2}\frac{\Lambda^4}{f^2} = (g f)^2 
\sim (1\ {\rm TeV})^2.
\end{equation} 
Therefore below 1 TeV, the matter content of the $T$-invariant minimal
moose model is simply the two-Higgs-doublet standard model, the same as 
the original $SU(3)$ minimal moose. At around 1
TeV, there are three sets of scalar triplets and singlets, two
doublets, as well as massive $SU(2)$ gauge bosons $B'$, $W'$ and $Z'$,
and one vector-like colored fermion responsible for
cancelling the quadratic divergence from the top loop.

\begin{figure}[tb]
\includegraphics[width=11cm]{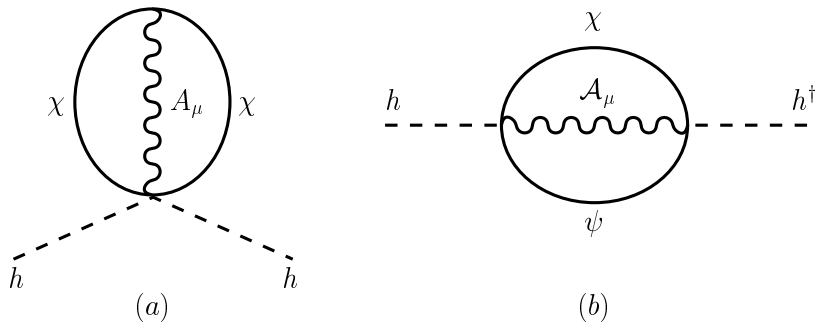}
\caption{\label{figuretwo}\it The diagrams responsible for cancelling the
  quartic divergence. Diagram $(a)$ is there only for the $U(1)_Y$ gauge
  boson since $\chi$ is an $SU(2)_W$ singlet and doesn't couple to $W$ and $Z$
  bosons directly. }
\end{figure}

Before moving on to other models, we note that for the minimal moose model
the quartic divergence from Fig.~\ref{figureone}, which gives the doublet a
heavy mass, is not problematic
because there are extra electroweak doublets in the model to spare.
For little Higgs models which do not contain any extra
electroweak doublet, it would give a contribution to the Higgs mass-squared
in the order of $(gf)^2\sim (1 \mbox{ TeV})^2$ if the cutoff is $\sim 10$
TeV, and hence destabilizing the electroweak scale. It is therefore
desirable to see if there is a way to cancel the two-loop quartic divergences. 
As we discussed above, the divergence
arises due to the incomplete fermion multiplet in the kinetic 
term~(\ref{fermionkinetic}) which breaks all chiral symmetries.
The remedy is
thus to complete the fermion multiplet by introducing 
a vector-like pair of Weyl fermions 
$\chi$ and
$\chi^c$ so that the fermion $\psi$ now transforms as a complete
fundamental under
the $SU(3)_V$:
\begin{equation}
\psi = \left( \begin{array}{c}
                     \psi_{D} \\
                       \chi        
               \end{array}        \right)
 \to U \psi.
\end{equation}
In this way, under the global $SU(3)_L\times SU(3)_R$ symmetry, $\xi \psi
\to L (\xi \psi)$ transforms like the fermion living on the $L$-site, while
$\xi^\dagger \psi \to R (\xi^\dagger \psi)$ transforms like the fermion
living on the $R$-site. Then each interaction in the kinetic term
Eq.~(\ref{fermion}) preserves either the $SU(3)_L$ or the $SU(3)_R$ which
keeps the doublet in $\xi_4$ from getting a mass. In terms of Feynman diagrams
the quartic divergence in Fig.~\ref{figureone} is cancelled by the
diagrams involving the $\chi$ fermion and the massive gauge bosons, as shown
in Fig.~\ref{figuretwo}. 
The fermions $\chi$ and $\chi^c$ are $T$-odd because $\psi \to \Omega_X \psi$ under $T$-parity. 
The Dirac mass $m_\chi \chi^c \chi$,
acting as the cutoff of the divergence in Fig~\ref{figureone}, needs to
be below $\sim 4\pi \sqrt{m_h f/g}$ for naturalness, which implies
$m_\chi$ only need to be as low as $\sim 3$ TeV.

In Ref.~\cite{Cheng:2003ju} a three-site model with $T$-parity was
constructed. The gauge couplings on the third site are large and the
particles associated with the third site are heavy, being close to the cutoff scale,
so it is appropriate to integrate them out below their masses. 
The low energy effective
theory after integrating out the third site is then described by the
$T$-invariant two-site model discussed here, except for the trivial 
choices of the
$SU(3)$ or $SO(5)$ groups. In this way, the three-site model may be viewed 
as a possible
UV extension of the two-site model, in which all fermions transform linearly
under the full symmetry, although in principle 
there could be other possibilities.

\section{An $SU(5)/SO(5)$ model with $T$ parity}
\label{sec:coset}

Having the experience with the QCD-like minimal moose model, we can
now generalize this construction to other little Higgs models. In
particular, we will focus on the littlest Higgs model 
based on an $SU(5)/SO(5)$ non-linear sigma model~\cite{Arkani-Hamed:2002qy}, 
which is the most
extensively studied little Higgs model in the literature.
It has been claimed that, in order to be compatible with the precision
data, the original model needs to be fine-tuned
and the viable parameter space is thus severely
constrained~\cite{Csaki:2002qg,Hewett:2002px,Csaki:2003si}. 
From the discussion in the previous sections, 
however, we can see that the strong
constraints can be removed completely, hence providing a natural
model for electroweak symmetry breaking if the $T$-parity can be
implemented. In this section we show that this is indeed possible. A
few more new states are needed to complete the construction, which will
be explained along the way.

Following the notation and the basis in
Ref.~\cite{Arkani-Hamed:2002qy}, we consider a non-linear sigma model
arising from a $5\times 5$ symmetric matrix $\Phi$, transforming under the
global $SU(5)$ symmetry as $\Phi \to V \Phi V^T$, with a vacuum
expectation value
\begin{equation}
\Sigma_0 = \left( \begin{array}{ccc}
                     &      & \openone \\
                     &   1  &          \\
           \openone  &      &     
                  \end{array}     \right),
\end{equation}
which breaks $SU(5) \to SO(5)$. The unbroken generators $T^a$ and broken
generators $X^a$ satisfy $\Sigma_0 T_a \Sigma_0 = - T_a^T$ and $\Sigma_0
X_a \Sigma_0 = X_a^T$, respectively. Same 
as $[SU(3)\times SU(3)]/SU(3)$, $SU(5)/SO(5)$ is also a symmetric space in which
the unbroken and broken generators satisfy the following schematic
commutation relations,
\begin{equation}
\label{eq:symmetric}
[T^a, T^b] \sim T^c, \quad [T^a, X^b] \sim X^c, \quad [X^a, X^b] \sim T^c.
\end{equation}
The Lie algebra of a symmetric space therefore has an automorphism: $T^a
\to T^a$ and $X^a \to -X^a$, which in the case of $SU(5)/SO(5)$ can be 
expressed as $\tau^a \to - \Sigma_0 {(\tau^a)}^T \Sigma_0$ for any
generator $\tau^a$. It is this automorphism that allows us
to define the $T$-parity consistently.
The Goldstone fields, $\Pi = \pi^a
X^a$, are parametrized as 
\begin{equation}
\Sigma(x) = e^{i\Pi/f} \Sigma_0 e^{i\Pi^T/f} = e^{i 2\Pi/f} \Sigma_0 .
\end{equation}
Two different $SU(2)\times U(1)$ subgroups of $SU(5)$ are gauged with equal strength:
\begin{eqnarray}
Q_1^a &=& \left( \begin{array}{cc}
            \sigma^a/2 &     \\
                       & \phantom{\sigma^a/2}
               \end{array}            \right),
 \quad \quad \ \ Y_1= \mbox{diag}(3,3,-2,-2,-2)/10 , \\
Q_2^a &=& \left( \begin{array}{cc}
            \phantom{\sigma^a/2} &     \\
                       & -\sigma^{a *}/2
               \end{array}            \right),
 \quad \,  \,\, Y_2= \mbox{diag}(2,2,2,-3,-3)/10 .
\end{eqnarray} 
The unbroken gauge group is identified with the electroweak $SU(2)_W\times
U(1)_Y$, under which the Goldstone bosons decompose into a doublet, taken
to be the
Higgs doublet, and a triplet:
\begin{equation}
\Pi = \left( \begin{array}{ccc}
            \phantom{\sigma} & \frac{H}{\sqrt{2}} & \phi \\
             \frac{H^\dagger}{\sqrt{2}}  &   & \frac{H^T}{\sqrt{2}}  \\
            \phi^\dagger     &  \frac{H^*}{\sqrt{2}}  &   
             \end{array}  \right),
\end{equation}
where we have omitted Goldstone bosons eaten by the Higgs mechanism. In
components the Higgs scalars are $H = (h^+\ h^0)^T$. A Higgs VEV,
$\langle H \rangle = (0\ v/\sqrt{2})^T$ triggers the electroweak
symmetry breaking. The
gauge interactions break the global $SU(5)$ symmetry, but they are chosen
in such a way that, when the gauge coupling 
of $[SU(2)\times U(1)]_1$ is turned off, there is an enhanced global
$SU(3)_1$ living in the upper-left $3\times 3$ block of the $SU(5)$
group. Conversely, when the coupling of $[SU(2)\times U(1)]_2$ is turned
off, there is an 
$SU(3)_2$ living in the lower-right $3\times 3$ block. In this way it is
guaranteed that the Higgs doublet does not receive one-loop quadratic
divergences from the gauge boson loops. 

The effective Lagrangian of the model is written as
\begin{equation}
{\cal L} ={\cal L}_K + {\cal L}_\psi + {\cal L}_t + {\cal L}_y ,
\end{equation}
where ${\cal L}_K$ contains the kinetic terms for the gauge and Goldstone
bosons; ${\cal L}_\psi$ includes the kinetic terms for the fermions; ${\cal
  L}_t$ generates the Yukawa coupling for the top quark; and ${\cal L}_y$
generates the Yukawa couplings for the light quarks.

The gauge and scalar kinetic terms ${\cal L}_K$ of the littlest Higgs model
is invariant under the $T$-parity which is defined as
\begin{eqnarray}
\Pi &\leftrightarrow& -\Omega\, \Pi\, \Omega,
\nonumber \\
A_1 & \leftrightarrow & A_2,
\end{eqnarray}
where $\Omega\equiv \mbox{diag}(1, 1, -1, 1, 1)$ commutes with all the
gauge generators and $\Sigma_0$. Similar to the previous section,
$\Omega$ is included in the definition of the $T$-parity to flip
the parity of the Higgs doublet. In this way, the light fields including
the standard model gauge fields and the Higgs doublet are even under
$T$-parity, while the heavy gauge bosons and the scalar triplet, which
picks up a mass $\sim 1$ TeV at one loop, are $T$-odd.

The kinetic term for the $SU(2)$ singlet fermions can be written down
as in Eq.~(\ref{eq:singlet}), then they only couple to the standard model
gauge bosons. 
To write down a kinetic term for the $SU(2)$ doublet fermion, we introduce
the object $\xi = \exp(i\Pi/f)$, just like in the $T$-invariant
minimal moose model. Since $\Sigma(x)=\xi^2 \Sigma_0$, the 
transformation property of $\xi$ under $SU(5)$ can be
deduced from that of $\Sigma(x) \to V \Sigma(x) V^T$:
\begin{equation}
\label{nonlinear}
\xi \to V \xi U^\dagger = U \xi (\Sigma_0 V^T \Sigma_0) 
\end{equation}
where $U=\exp(i u^a(\Pi,V) T^a)$ belongs to the unbroken $SO(5)$ and
is a non-linear representation of the $SU(5)$; the function $u^a$
depends on the Goldstone fields $\Pi$ and the $SU(5)$ rotation $V$ in a
non-linear way. For a set of fermions furnishing a complete multiplet of
the unbroken $SO(5)$, the kinetic term is written down as in
Eq.~(\ref{fermion}). Here we just need to know
how to embed an electroweak doublet in an $SO(5)$ representation.
Since it is easy to embed an electroweak doublet in an $SU(5)$
representation, we begin by observing that, in the basis we choose, 
a fundamental representation of
$SU(5)$ decomposes into two copies of fundamental representation of 
the unbroken $SO(5)$:
\begin{equation}
 \psi= \left( \begin{array}{c}
          \psi_1  \\
          \chi    \\
          \psi_2 
        \end{array}  \right)  \to  \psi \pm \Sigma_0 \psi^*,
\end{equation}
where $\psi_1$, $\psi_2$ are doublets and $\chi$ is a singlet under $SU(2)_V$.
Now we can simply put an electroweak doublet fermion $\psi_D$ of the 
standard model in the position of $\psi_1$. However, without any additional
fermions, the kinetic term breaks all global symmetries that protect
the Higgs mass, and a two-loop quartic-divergent contribution to the 
Higgs mass-squared will be generated as discussed in the previous section.
There are two ways to deal with this problem. The first is to enlarge the
scalar sector and introduce additional electroweak doublet, just like in
the moose model. For example, we can double the Goldstone bosons by
having two $\Sigma$ fields. Then there are two electroweak doublets. 
One doublet
can get a TeV mass from the fermion kinetic terms by the two-loop quartic
divergences, but the other can remain light and be the SM Higgs doublet.
Other additional Goldstone bosons such as triplets and singlets can be lifted
by either the one-loop contributions or appropriate ($\Omega$-dependent) 
plaquette operators.

Alternatively,
we can cancel the quartic divergences by introducing new
vector-pair of singlet and doublet fermions, $\chi$, $\chi^c$
and $\tilde{\psi}_D$, $\tilde{\psi}^c_D$, respectively, 
for each electroweak doublet
fermion in the standard model. Put $\psi_D$, $\chi$, $\tilde{\psi}_D$ 
into the complete multiplet of the $SO(5)$:
\begin{equation}
\label{partial}
\psi= \left( \begin{array}{c}
          \psi_D  \\
          \chi    \\
          \tilde{\psi}_D
        \end{array}  \right)  \to U \psi
\end{equation}
and introduce Dirac mass terms, $m_\chi \chi^c \chi$ and $m_{\tilde{\psi}_D}
\tilde{\psi}_D \tilde{\psi}^c_D$, for the new fermions. 
The kinetic term is now written as
\begin{equation}
\ov{\psi}\, \bar{\sigma}^\mu  \left( \partial_\mu + v_\mu^a T^a
+ i g_Y (Y_{\psi_D} -Y_H)B_\mu \right) \psi,
\label{cosetfermion}
\end{equation}
where $v_\mu^a$ is defined as in Eq.~(\ref{oneform}).
The kinetic term is invariant under the $T$-parity with
$\psi$ transforming as $\psi \to \Omega \psi$, which implies that $\chi$,
$\chi^c$ are $T$-odd and $\tilde{\psi}_D$, $\tilde{\psi}^c_D$ are $T$-even. 
The kinetic term does not contain any direct couplings
between standard model fermions and the heavy gauge bosons.  
They are contained in the $T$-odd term
\begin{equation}
\ov{\psi}\, \bar{\sigma}^\mu p_\mu^a X^a \psi,
\end{equation}
which is forbidden by $T$-parity.

There is still a question about how to embed the doublet $\tilde{\psi}^c_D$
which marries with $\tilde{\psi}_D$. In the low energy theory, a possibility
is to embed $\tilde{\psi}^c_D$, $\chi^c$, together with an additional singlet
$\chi^{\prime c}$ into a spinor representation of $SO(5)$. ($\chi^{\prime c}$
can get a mass $m_{\chi'} \chi^{\prime c} \chi^{\prime}$ with another singlet
$\chi^{\prime}$.) Then its kinetic term can be written down following
CCWZ. Alternatively, one can extend the theory with a third $SU(2)$ gauge
group, which is neutral under the $T$-parity, and have $\tilde{\psi}^c_D$
be a doublet of this extra $SU(2)$. This extra $SU(2)$ is broken together
with the two $SU(2)$ gauge groups inside the $SU(5)$ down to the diagonal
SM $SU(2)$. The Dirac mass term $m_{\tilde{\psi}_D} \tilde{\psi}_D
\tilde{\psi}^c_D$ can then arise from the Yukawa type interactions
after the breaking. The gauge coupling of the third $SU(2)$ can be large so 
that the mass of the extra gauge bosons is near the cutoff, and the SM gauge 
fields are mostly made of the gauge fields of the two $SU(2)$'s inside
the $SU(5)$. This is in the similar spirit as the three-site moose model
of Ref.~\cite{Cheng:2003ju}. In fact, such extensions also allow us to 
construct UV 
extensions of the $T$-invariant $SU(5)/SO(5)$ model, which contain only linear 
representations, {i.e.,} no CCWZ construction is needed~\cite{threesite}. 
More generally, a little Higgs model based on a symmetric space $G/H$ can 
be made $T$-invariant by extending it to $(G\times H)/H$. In this way, all
fermions can be assigned 
to linear representations under the full group $G\times H$ without invoking
CCWZ.


It can be explicitly checked that, in the presence of $\chi$ and 
$\tilde{\psi}_D$, the
quartic divergence in Fig.~1 is cancelled by diagrams
involving $\chi$ and $\tilde{\psi}_D$ similar to those in Fig.~2.
Because we have the complete fermion multiplet, the only symmetry breaking
sources in the kinetic term come from the gauge generators $Q^a_{1,2}$.
We already know that each gauge generator 
preserves either $SU(3)_1$ or $SU(3)_2$ in the global $SU(5)$ group,
so for each interaction
\begin{equation} 
\ov{\psi}\, \xi^\dagger\, Q^a_{1,2}\, \xi\, \psi  \quad \quad \mbox{and}
\quad \quad \ov{\psi}\, \xi\, Q^a_{1,2}\, \xi^\dagger\, \psi ,
\end{equation}
there is an unbroken
$SU(3)$ symmetry protecting the Higgs doublet from getting a mass.
A Higgs mass can be induced only in the presence of both $Q^a_1$
and $Q^a_2$, which requires going to higher loops or a mass insertion
between $A_1^a$ and $A_2^a$. Either way the contribution to the Higgs
mass-squared can be no bigger than the two loop {\it quadratic} divergences and
hence is safe.  
Because the quartic divergence in Fig.~1 is cut off by the masses of the extra
fermions, $m_\chi$, $m_{\tilde{\psi}_D}$, for naturalness they need to
be below $\sim 4\pi \sqrt{m_h f/g}$. 


Next step is to write down a top Yukawa coupling without introducing the
associated quadratically divergent contribution to the Higgs mass, which can be achieved by 
adding the usual pair of $T$-even colored fermions ${t'}$ and ${t'}^c$, as well as
three additional colored, fermionic weak singlet fermions, $u^c, u^c_T$, and $U$.
Grouping the third generation quark doublet $q_3=(b\, t)$ 
with ${t'}$ and $\tilde{q}_{3}$ into a row
vector $\tilde{Q}_3^T = ( q_3\ {t'}\ \tilde{q}_{3})$, the top Yukawa coupling
${\cal L}_t$ can be written as\footnote{We thank J. R. Espinosa for
  pointing out the Yukawa coupling in an early version of this paper leads
  to unacceptably large Higgs mass.}
\begin{eqnarray}
{\cal L}_t &=& \frac{1}{4\sqrt{2}}\lambda_1 f \epsilon_{ijk} \epsilon_{xy}\left [ 
   (\xi \tilde{Q}_3)_i (\xi^2 \Sigma_0)_{jx} (\xi^2 \Sigma_0)_{ky} u^c
   + (\tilde{\xi} \tilde{Q}_3)_i (\tilde{\xi}^2 \Sigma_0)_{jx}
   (\tilde{\xi}^2 \Sigma_0)_{ky} u^{c}_T\right] \nonumber \\
 &&  + \lambda_2 f {t'}{t'}^c + 
   \frac{\lambda_3 f}{\sqrt{2}} U (u^c-u^{c}_T) + \mbox{h.c.} ,
\label{topyukawa}
\end{eqnarray} 
where $i,j,k=1,2,3$, $x,y=4,5$, and $\tilde{\xi}=\Omega \xi^\dagger
\Omega$. This is nothing but the $T$-symmetrized version of 
the top Yukawa interaction written down
in Ref.~\cite{Arkani-Hamed:2002qy},  
which can be easily recognized by noting that $\xi^2 \Sigma_0 =
\Sigma(x)$. We define $u^{c}_T$ to be the image of $u^c$ under
$T$-parity, so that
the first term in the square bracket in Eq.~(\ref{topyukawa}) is mapped into
the second under the $T$-parity. The singlet fermion $U$ is taken to be
$T$-odd and marries the $T$-odd linear combination $U^c\equiv (u^c - u^c_T)/\sqrt{2}$.
Any $T$-odd interactions have been projected out in Eq.~(\ref{topyukawa}).
One way to show that there are no one-loop
quadratic divergences generated from Eq.~(\ref{topyukawa}) is to perform a
spurion analysis as in Ref.~\cite{Arkani-Hamed:2002qx}. For that purpose we
can regard ${\cal L}_t$ as having 
four independent interactions with coupling constants $\lambda_1, \lambda_1^\prime,
\lambda_2$ and $\lambda_3$, where $\lambda_1^\prime$ is the coefficient of
$(\xi \tilde{Q}_3)_i (\xi^2 \Sigma_0)_{jx} (\xi^2 \Sigma_0)_{ky} u^c_T$ and
forced by $T$-parity to be equal to $\lambda_1$. Each interaction in ${\cal
L}_t$ preserves enough $SU(3)$ global symmetries to individually protect
the Higgs mass from getting one-loop quadratic divergences; it takes more
than one $\lambda$ to get a quadratically divergent mass. Moreover, with
the introduction of new fermions, each
$\lambda$ has a global $U(1)$ rephasing symmetry associated with
redefining the phases of $u^c, u^c_T, t^{\prime c}$ and $U$. Therefore these
spurions can only enter as $|\lambda^{(\prime)}/4\pi|^2$ and the quadratic
divergence does not come in until two-loop order. Without the rephasing
symmetries, there could be divergences proportional to, for instance,
$(\lambda_1/4\pi)(\lambda_1^\prime/4\pi)$ which would be too big a
contribution to the Higgs mass-squared.

Expanding ${\cal L}_t$ to leading order in the Higgs
particle gives
\begin{equation}
 \lambda_1 (\sqrt{2} H^\dagger q_3 + f {t'}) u_3^c + \lambda_2 f
 {t'}{t'}^c + \cdots,
\end{equation}
where $u_3^c=(u^c+u^c_T)/\sqrt{2}$ is the $T$-even linear combination of
$u^c$ and $u_T^c$. As usual, a linear combination $T^c \approx (\lambda_1 u_3^{c}+\lambda_2 {t'}^c)/
\sqrt{\lambda_1^2+\lambda_2^2}$ marries with $T \approx t'$ 
(plus a small mixture from
$t$) and becomes heavy with a mass $M_T \approx 
\sqrt{\lambda_1^2+\lambda_2^2} f$.
After integrating out the heavy top partner $T$, 
the orthogonal combination $t^c$ has the desired
Yukawa coupling to $q_3$ with the strength $\lambda_t = \sqrt{2}\lambda_1
\lambda_2/\sqrt{\lambda_1^2+ \lambda_2^2}$.

\section{Little Higgs phenomenology with $T$-parity}
\label{sec:pheno}

Little Higgs models without $T$-parity are in general strongly constrained
by the precision electroweak measurements. The leading corrections to
the electroweak observables in these models come from several sources.
First, the heavy gauge bosons mix with the SM gauge bosons through
vertices like $H^\dagger W^\mu W_\mu^\prime H$ after electroweak symmetry
breaking, when the Higgs doublet gets a VEV. This shifts the masses of
$W$ and $Z$ and also modifies the couplings of $W$ and $Z$ to the SM fermions.
Secondly, integrating out heavy gauge bosons gives additional contributions to 
4-fermion 
interactions at low energies if the heavy gauge bosons couple to SM 
fermions directly. In particular, the contribution to the Fermi constant
$G_F$ from muon decays feeds into all precision measurements since $G_F$ is
used as an input parameter. Finally, for models containing scalar triplets,
a VEV of the triplet $\langle \phi\rangle$ will be induced by the Higgs VEV
if there is a coupling $H \phi H$. The triplet VEV also shifts the $W$ 
boson mass. These corrections are all of the order $(\langle H
\rangle /f)^2$. Because many electroweak observables such as $W$ mass and
$Z$ couplings to fermions are very accurately measured, 
the scale of $f$ is in general constrained to be  somewhat larger than 1 TeV
as most phenomenological studies on little Higgs theories found. This in turn
re-introduces fine-tuning in the Higgs mass. For example, in the original
littlest Higgs model, the $SU(5)/SO(5)$ model without $T$-parity, the scale
$f$ needs to be above 4 TeV or so \cite{Hewett:2002px,Csaki:2002qg,Csaki:2003si}
in order to be compatible with precision data.

On the other hand, with $T$-parity all the above tree-level corrections
to the electroweak observables are absent because they violate the
$T$-parity. Apart from the unknown contributions from the cutoff physics,
the corrections to the electroweak observables within the
effective theory are loop-suppressed. One example of such constraints is
the correction to $\Delta \rho$ parameter due to the loop of the 
heavy top quark
partner, which is responsible for cancelling the quadratic divergence
from the top Yukawa coupling. This correction is in the order of  $(\langle H
\rangle /f)^2/16\pi^2$ and very small.
The correction to the $Z\bar{b}b$ vertex also gives comparable effect. In
addition, the heavy gauge bosons could also contribute to the $\rho$
parameter at one loop level. All these constraints are down by a loop
factor and allow the scale $f$ to be well below 1 TeV.

$T$-parity also drastically affects the collider phenomenology of the little
Higgs theories.
The $T$-odd particles
cannot be singly produced due to $T$-parity, which implies that 
direct searches must rely on pair-productions. When
the $T$-odd particles are produced, they will eventually decay into the LTP,
the lightest $T$-odd particle, which is stable since $T$-parity forbids it
{}from decaying into lighter particles which are all $T$-even. A charged LTP is
not preferred as it can cause cosmological 
problems. In $T$-invariant models the LTP is most likely to be the $B^\prime$
gauge boson because its mass is suppressed by the smallness of the $U(1)_Y$ 
coupling and also often by the
normalization factor of the $U(1)$ generator, although
some other neutral scalar is also possible. For a neutral LTP, the typical
collider signals will be the 
$T$-odd particles decaying into jets and/or leptons plus missing energies. 
In this respect
the phenomenology is similar to those of the supersymmetric theories with
$R$-parity and KK-parity conserving universal extra-dimensions 
(UED)~\cite{Appelquist:2000nn,Cheng:2002iz,Cheng:2002ab}, where
the lightest supersymmetric particle (LSP) and the lightest KK-odd particle
(LKP) are also stable and would escape detection in collider experiments if
they are neutral. Moreover, the LTP could serve as a viable dark matter
candidate, similar to LSP and LKP. 
One distinct feature of the $T$-invariant little Higgs theories is that 
the top sector still contains a $T$-even heavy top quark partner,
whose phenomenology is not affected by the $T$-parity. It can be used
to test the little Higgs mechanism in colliders, as shown in Ref.~\cite{Perelstein:2003wd}.
The $T$-even top partner can be singly
produced and its decay products do not necessarily give missing energies.
A recent study show that it can be observed up to mass of order 2.5 TeV
via its decay to $Wb$ at the LHC~\cite{Azuelos:2004dm}.
On the other hand, for the $T$-odd particles such as heavy gauge bosons
and scalar triplets, all the previous phenomenological studies without
assuming the $T$-parity do not apply. The details of their 
phenomenology depend on the particle content and the spectrum of each
individual model. In the rest of this section, we will describe in a
little more detail the phenomenology of the $T$-invariant
$SU(5)/SO(5)$ model as an example.

Compared with the original littlest Higgs model, the $T$-invariant model
has an extended scalar sector or fermion sector (or both) in the TeV scale.
Here we consider the model with an extended fermion sector. 
Then, in the scalar and gauge sectors, the $T$-invariant $SU(5)/SO(5)$ model 
has
the same matter/field content as the littlest Higgs model. There is one Higgs
doublet $H$ which is light, one heavy triplet scalar $\phi$, and a set of massive
$SU(2)\times U(1)$ gauge bosons $W^\prime$ and $B^\prime$.  The triplet
scalar, responsible for cancelling
the quadratic divergence of the Higgs quartic coupling, cannot get a VEV
here because it is $T$-odd and forbidden to have a tadpole term from
coupling to the Higgs doublet after electroweak symmetry 
breaking.  The gauge couplings of the two
$SU(2)\times U(1)$ groups are required to be equal by $T$-parity, implying $g_1=g_2=\sqrt{2}g$
and $g_1^\prime=g_2^\prime=\sqrt{2}g_Y$, where $g$ and $g_Y$ are the gauge
coupling constants for the electroweak $SU(2)_W\times U(1)_Y$. The mixing
angles in the gauge couplings are set by the $T$-parity to be $\pi/4$, instead of being
free parameters as in the littlest Higgs model. The masses for
various gauge bosons are given by 
\begin{eqnarray}
M_W^2 &=& \frac12 g^2 f^2 \sin^2\frac{\langle H \rangle}{f}, \nonumber \\
M_Z^2 &=& \frac12 (g^2+g_Y^2) f^2 \sin^2\frac{\langle H \rangle}{f}, 
\nonumber \\
M_{W'}^2 &=& g^2 f^2 -\frac12 g^2 f^2 \sin^2\frac{\langle H
  \rangle}{f},
\nonumber \\
\left( \begin{array}{ll} M_{B'}^2 & M_{B' W'_3}^2 
\\ M_{B' W'_3}^2 & M_{W'_3}^2 \end{array} \right)
&=&
\left( \begin{array}{cc} \frac{1}{5}g_Y^2 f^2-\frac{1}{8}g_Y^2 f^2
\sin^2\frac{2 \langle H \rangle }{f} & 
\frac{1}{8} g g_Y f^2 \sin^2\frac{2 \langle H \rangle}{f} \\
\frac{1}{8} g g_Y f^2 \sin^2\frac{2\langle H \rangle}{f} &
g^2 f^2 -\frac{1}{8}g^2 f^2 \sin^2\frac{2 \langle H \rangle}{f}
\end{array} \right) .
\label{exact_mass}
\end{eqnarray}
In the limit $f \gg \langle H \rangle$, these formulae reduce to the
familiar ones:
\begin{equation}
\label{wprimemass}
M_W^2 \approx \frac14 g^2 v^2, \quad M_Z^2 \approx \frac14(g^2+ g_Y^2)v^2,
  \quad M^2_{W^\prime} \approx g^2 f^2, \quad M^2_{B^\prime} 
  \approx \frac15 g_Y^2 f^2 .
\end{equation}

In the fermion sector, in addition to the top partners which
cancels the quadratic divergence from the top Yukawa coupling, 
for every SM doublet fermion there
is also a vector-like $T$-odd $SU(2)_W$ singlet fermion and a vector-like
$T$-even doublet fermion. 
If the masses of these fermions are not degenerate among different generations
or aligned with the SM fermion masses, they can induce flavor-changing
effects, such as $K^0-\overline{K}^0$ mixing, $\mu \to e\gamma$ and so on.
However, their couplings to SM fermions always involve the Higgs or
the scalar triplet,
{\it e.g.}, $(1/f)\,\bar{\chi} \bar{\sigma}^\mu {\cal A}_\mu H^T \psi_{\rm SM}$,
$(1/f^2)\,\bar{\tilde{\psi}} \bar{\sigma}^\mu A_\mu H H^T \psi_{\rm SM}$,
$ \bar{\tilde{\psi}} \bar{\sigma}^\mu {\cal A}_\mu \phi\, \psi_{\rm SM}$.
Their contributions to the flavor-changing neutral currents are therefore
suppressed either by powers of $\langle H \rangle /f$, 
or by higher loops if the physical Higgs boson or the scalar triplet
is involved. For $f,\, 
m_{\chi, \tilde{\psi}} \sim$ 1 TeV, the induced flavor-changing effects
are already safe even without degeneracies or alignments.

As we discussed earlier in this section, within the effective  theory the
corrections to the electroweak observables are loop-suppressed.
The strongest constraint comes from the $\Delta \rho$ correction
due to heavy gauge boson loops. One can see from Eq.(\ref{exact_mass}) that
even if $g_Y$ is turned off, the heavy $W^{\prime \pm}$ and $W'_3$ gauge bosons
are not degenerate and the mass splitting is 
\begin{equation}
\Delta M^{\prime 2} = M^2_{W^\prime_3} - M^2_{W^{\prime \pm}} = \frac12 g^2 f^2
  \sin^4 \frac{\langle H \rangle}{f}.
\end{equation}
This mass splitting induces at one-loop level a custodial $SU(2)_C$
breaking effect and the correction to $\Delta \rho$ is given by
\begin{equation}
\label{rhow}
\Delta \rho_{W^\prime} = -\frac{e^2}{s_W^2 c_W^2 M_Z^2}\left( \frac{9 g^2}{64\pi^2} \Delta
  M^{\prime 2} \log \frac{\Lambda^2}{M_{W^\prime}^2} \right) ,
\end{equation}
where $s_W^2 = 1- c_W^2 = \sin^2 \theta_W$, and $\theta_W$ 
is the Weinberg angle. 
The logarithmic divergence means that a counter term 
in the non-linear sigma model of the form
\begin{equation}
\frac{g^2}{16\pi^2} \beta f^2 \sum_{i,a} \mbox{Tr}[(Q_i^a D_\mu \Sigma)(Q_i^a
  D^\mu \Sigma)^*] 
\end{equation}
is required to cancel the divergence,
where the numerical coefficient $\beta$ is expected to be of order unity. 
Indeed, after
turning on the Higgs VEV, this interaction 
gives a mass-splitting within the standard model $SU(2)$ gauge bosons
\begin{equation}
\Delta M^2 = M_{W_3}^2 - M_{W^{\pm}}^2 = \beta \frac{g^4 f^2}{64\pi^2} \sin^4
\frac{\langle H \rangle}f.
\end{equation} 
This is due to the fact that gauging two $SU(2)$'s in this model is
a custodial $SU(2)_C$ violating effect. 
There are further $SU(2)_C$ breaking effects from the $U(1)$ gauge couplings, 
but they are expected to be smaller due to the smallness of $g_Y$. 
Assuming there is no large counter term at the cutoff,
the contribution to $\Delta \rho$ from $W'$ loop between the $M_W'$
and the cutoff $\Lambda$ is negative and enhanced by a $\log (\Lambda/M_{W'})$
factor,
\begin{equation}
\Delta \rho_{W^\prime} \approx -0.002 \left( \frac{450\ \mbox{GeV}}f \right)^2 \left( 1+ 0.34
\log \frac{\Lambda}{4\pi f} \right).
\end{equation}
If one requires that $\Delta \rho > -0.002$, this only constrains 
the symmetry breaking scale $f$ to be greater than $450$ GeV which corresponds
to  
\begin{equation}
M_{W^\prime ,Z^\prime} \gtrsim 280\, \mbox{GeV}, \quad \quad 
M_{\hat{B}^\prime} \gtrsim 60 \, \mbox{GeV}, \quad \quad
M_T \gtrsim 640 \, \mbox{GeV},
\end{equation}
where $\hat{B}^\prime$, $Z'$ represent the mass eigenstates and are very
close to the gauge eigenstates $B'$, $W'_3$.

The $\Delta \rho$ parameter also receives contribution from 
the top partner at the one-loop level,
\begin{equation}
\label{toprho}
\Delta \rho_{t'} \approx \frac{3\lambda_t^2}{16\pi^2} \sin^2\theta_t 
\left[
\ln\left(\frac{M_T^2}{m_t^2}\right) -1+\frac{1}{2} 
\left(\frac{\lambda_1}{\lambda_2}\right)^2 \right],
\end{equation}
where
\begin{equation}
\sin^2\theta_t \approx \left(\frac{\lambda_1}{\lambda_2}\right)^2
\frac{m_t^2}{M_T^2}
\end{equation}
is the mixing angle between the top quark and the top partner in the
left-handed sector, and $M_T \approx \sqrt{\lambda_1^2 +\lambda_2^2}f$ is the mass of the
heavy top partner. The top Yukawa coupling is given by
$\lambda_t \approx {\sqrt{2}\lambda_1 \lambda_2}/{\sqrt{\lambda_1^2
+\lambda_2^2}}$. 
It does not provide any significant constraint as long as $\lambda_1/\lambda_2$
is somewhat small ($\lesssim 0.8$). 
The correction to the $Z\bar{b}b$ vertex from the top partner loop has
the same leading logarithm as in Eq.~(\ref{toprho}) and gives comparable
constraint to $\Delta \rho_{t'}$.
The contributions to the $S$ parameter of 
Peskin-Takeuchi~\cite{Peskin:1990zt,Peskin:1991sw} are also small. 
In fact, the top partner contribution to $\Delta \rho$ is positive
so it is possible to partially cancel the negative contribution from the 
$W'$ loop, allowing $f$ to be even lower. (The direct search limit is 
$m_{W'} \gtrsim 100$ GeV from LEP II which corresponds to $f \approx 200$ GeV.)
Of course, a very low $f$ undermines the purpose of the little Higgs theories
and one expects that the higher dimensional operators at the cutoff 
$\Lambda=4\pi f$ will give too big contributions to the 
electroweak observables. 
Indeed, as the constraints from the calculable effects within the effective
theory are rather weak in this model, the most stringent constraints 
are expected
to come from the higher dimensional operators
arising from unknown UV physics at the cutoff scale, 
which would prefer $\Lambda \approx 4\pi f \gtrsim 5-10$ TeV
if these unknown contributions are not suppressed.

In terms of collider phenomenology, the $T$-invariant model escapes
conclusions reached by most studies in the literature so far, except for
those on the $T$-even top partners $(T,T^c)$. The $T$-odd particles in the model are the
heavy gauge bosons, a triplet scalar, and the singlet fermions
$\chi_{q_i},\,\chi_{l_i},\,i=1,2,3$ and $(U, U^c)$; 
they are produced in pairs through the SM gauge bosons
and eventually decay into the LTP. 
It can be seen
from Eq.~(\ref{wprimemass}) 
that the gauge bosons are likely to be lighter than the triplet scalar and
the heavy fermions.
The relatively light $B^\prime$ boson is most likely to be the
LTP, as will be assumed in the following studies. 
Moreover, electroweak symmetry breaking induces
corrections to the mass of the heavy gauge bosons, as well as a small mixing
between $W_3^\prime$ and $B^\prime$, which causes the mass eigenstate
$Z^\prime$ to be slightly heavier than the $W^\prime$. The mass splitting
can be calculated from Eq.(\ref{exact_mass}) and 
is of the order of $\langle H \rangle^4/f^2$. The triplet scalar
$\phi$ can be slightly heavier than 1 TeV, since it is
only responsible for cancelling the quadratic divergence from the Higgs
quartic coupling. The singlet fermions $\chi$ and doublet fermions
$\tilde{\psi}$ can be 3 TeV or higher, because they are there only to
cancel the quartic divergence from the fermion kinetic term. The additional
fermions in the top sector, on the other hand, is at around 1 TeV or so.

Due to the $T$-parity, all the $T$-odd
particles need to decay into another $T$-odd particle plus something, and
will eventually cascade down to the LTP, resulting in the missing energy
in the detector. $W'$ is likely to be the next-to-lightest $T$-odd particle.
It decays into $W$ plus $\hat{B}^\prime$ through the small 
$W_3^\prime$ component
in $\hat{B}'$ due to the mixing. 
$Z'$ decays into $W\, W\, \hat{B}'$ through 
$W\, W^{\prime *}$
(one of them needs to be virtual because of small mass splitting between
$W'$ and $Z'$).
Other heavy particles can decay through the interactions,
$\phi^\dagger {\cal A}^\mu D_\mu (H\,H^T)$, $\bar{\chi} \bar{\sigma}^\mu 
{\cal A}_\mu H^T \psi_{\rm SM}$, $\bar{\tilde{\psi}} \bar{\sigma}^\mu 
{\cal A}_\mu H \chi$,
$\bar{\tilde{\psi}} \gamma^\mu A_\mu H H^T \psi_{\rm SM}$,
$ \bar{\tilde{\psi}} \gamma^\mu {\cal A}_\mu \phi\, \psi_{\rm SM}$, and
$H^\dagger \phi q_3 U^c$,
where $H$ can be replaced by its VEV or the physical
Higgs boson. 

\begin{table}[t]
\caption{\label{decay}\it The major decay channels of the heavy particles 
in the $T$-invariant $SU(5)/SO(5)$ model. The $B^\prime$ is assumed to be 
the LTP. The particles produced in the decays may be replaced by the ones
in the parentheses.
}
\begin{tabular}{|c|c|c|} \hline
\phantom{aaa} Particle \phantom{aaa} & \phantom{aaa}  $T$-parity
\phantom{aaa} & \phantom{aaa} Major decay channels\phantom{aaa} \\ \hline 
$W^\prime$ & $-$         & $W \, \hat{B}^\prime$  \\ \hline
$Z^\prime$ & $-$         &  $W\, W\, \hat{B}^{\prime}$ \\ \hline
$ \phi  $  & $-$         &  $W(Z,h)\, W^\prime (Z', \hat{B}') $  \\ \hline
$ t^\prime$  & +  &      $t\,h,\, t\,Z,\, b\,W$ \\ \hline
$ U $   &  $-$  &   $ t \, \phi   $ \\ \hline
$ \chi$  & $-$         &  $\psi_{\rm SM}\, W^\prime (Z', \hat{B}')$ 
\\ \hline
$ \tilde{\psi}$ & +  &   $\chi \, W^\prime (Z', \hat{B}'), \, 
\psi_{\rm SM}\, W (Z) $
\\ \hline
\end{tabular}
\end{table}

In Table~\ref{decay} we list the major decay modes for all the heavy
particles in the $T$-invariant model. 
For concreteness we have assumed the following spectrum,
\begin{equation}
M_{\hat{B}'} < M_{W'} < M_{\phi} < m_T \lesssim m_U < m_{\chi} \lesssim m_{\tilde{\psi}}. 
\end{equation}
The decays of the lighter (and hence more accessible) 
states ($W',\, Z',\,
\phi$) often produce $W$, $Z$ bosons associated with the missing energy,
which could be the first things to look for in collider experiments,
though separating them from SM backgrounds would be a challenge.

\section{Conclusions}
\label{sec:con}

With Tevatron Run II currently running and LHC to start in 2007, the TeV
scale physics will be fully explored in the coming decade. The mystery
of the origin of the electroweak symmetry breaking is expected to be unveiled.
So far the precision electroweak measurements have not provided any evidence
for new physics beyond the standard model, which is quite puzzling.
Perhaps the fact that there is no sign of new physics in the precision electroweak
data itself is one of the biggest hints on the possible new physics that
will show up at the TeV scale. In this paper, we show that the little
Higgs theories, supplemented with the $T$-parity, can be a solution to this
little hierarchy problem. The large quadratically divergent corrections
to the Higgs mass-squared from the standard model particles are cancelled
by those from new TeV particles, which stabilizes the
electroweak scale. At the same time 
the $T$-parity forbids the tree-level contributions
to the electroweak observables from the TeV particles, and makes the little
Higgs theories
consistent with the electroweak precision data without fine-tuning.

A subtlety for imposing $T$-parity on little Higgs theories is
how to incorporate the standard model fermions. 
However, here we show that by following
CCWZ, which is the most general way to construct the low energy effective 
theory for a broken symmetry, this problem is readily resolved. The essential
$T$-even interactions and the troublesome $T$-odd interactions can be
naturally separated, which allows for an obvious implementation of $T$-parity.
Although we only discussed two examples, the minimal moose model and the
littlest Higgs model, from the discussion it should be apparent that the
$T$-parity can be introduced in many other models. As long as a model is
based on a symmetric space with the unbroken and broken generators
satisfying (\ref{eq:symmetric}), the automorphism $T^a\to T^a$, $X^a\to -X^a$,
allows us to define a $T$-parity consistently. For some models one needs to 
worry about the 2-loop quartic-divergent contributions 
to the Higgs mass-squared,
but it can be easily addressed by completing fermions into complete
multiplets of the unbroken group $H$ or enlarging the global symmetry.

There are also little Higgs models which do not live in a symmetric coset space.
One example is the little Higgs model based on a ``simple group'' 
\cite{Kaplan:2003uc}. In the simplest version (which all other variants 
based on), an $SU(3)$ gauge symmetry is broken down to $SU(2)$ by the VEVs
of two triplets. In this case, the broken generator $T^8$ does not
satisfy (\ref{eq:symmetric}), and the coset space is not a symmetric space. One
cannot find a consistent definition of $T$-parity under which all heavy
gauge bosons are odd. Nevertheless, one can still follow the approach
of CCWZ. The model-independent couplings of the SM fermions to
the SM gauge bosons and the model-dependent couplings of the SM fermions to the
heavy gauge bosons are still naturally separated into different terms.
The difference is that in this case there is no symmetry to forbid the 
couplings to the heavy gauge bosons. In fact, they will be radiatively
generated even if one does not include them in the first place. For
such models, the electroweak precision constraints turn into constraints
on these model dependent couplings. How much fine-tuning, if any, 
is required depends on the model and needs to be examined individually.

The existence of an exact $T$-parity obviously has a big impact on
the little Higgs phenomenology. The $T$-odd particles need to be 
pair-produced and the lightest $T$-odd particle is stable. A neutral
LTP provides a viable dark matter candidate and gives rise to missing
energy signals in collider experiments. To a first approximation, the
collider signals for this class of models are  similar to those of
the supersymmetric standard model. The details of what can be observed at
the colliders depend on each individual model and its spectrum. They certainly
deserve more thorough investigations. 
In order to distinguish the little Higgs theories with $T$-parity
from supersymmetry, the information on the spins of the new
particles participating in the cancellation of quadratic divergences will be
crucial. In this respect, a linear collider with a high enough energy 
could be very helpful. 

\begin{acknowledgments}

We would like to thank Nima Arkani-Hamed who provided many important
insights to this work.
We would also like to acknowledge helpful conversations with Jose Espinosa, Markus Luty,
Jay Wacker and Martin Schmaltz. This work is supported in part by the National Science Foundation under
grant PHY-0244821.
\end{acknowledgments}


\end{document}